\documentclass[preprint]{imsart}

\RequirePackage[OT1]{fontenc}
\RequirePackage{amsthm,amsmath,amssymb}
\RequirePackage{natbib}
\RequirePackage[colorlinks,citecolor=blue,urlcolor=blue]{hyperref}
\usepackage{graphicx}

\graphicspath{{./Images/}}

\startlocaldefs
\numberwithin{equation}{section}
\theoremstyle{plain}

\endlocaldefs
\begin{document}

%
%
%
%

\begin{frontmatter}
	\title{Extending Bayesian structural time-series estimates of causal impact to many-household conservation initiatives}
	\runtitle{Extending BSTS impact estimates to conservation programs}
	
	\begin{aug}
		\author{\fnms{Eric} \snm{Schmitt}\ead[label=e1]{eric.schmitt@protix.eu}},
		\author{\fnms{Christopher} \snm{Tull}\ead[label=e2]{chris@argolabs.org}}
		\and
		\author{\fnms{Patrick} \snm{Atwater}
			\ead[label=e3]{patrick@argolabs.org}}


		\address{Address of the First author\\
			Protix\\
			Industriestraat 3\\
			5107 NC, Dongen, The Netherlands\\
			\printead{e1}}

				\address{Address of the Second and Third authors\\
			California Data Collaborative\\
			525 S Hewitt St, Los Angeles, CA 90013}
		
	\end{aug}

\begin{abstract}

Government agencies offer economic incentives to citizens for conservation actions, such as rebates for installing efficient appliances and compensation for modifications to homes. The intention of these conservation actions is frequently to reduce the consumption of a utility. Measuring the conservation impact of incentives is important for guiding policy, but doing so is technically difficult. However, the methods for estimating the impact of public outreach efforts have seen substantial developments in marketing to consumers in recent years as marketers seek to substantiate the value of their services. One such method uses Bayesian Stuctural Time Series (BSTS) to compare a market exposed to an advertising campaign with control markets identified through a matching procedure. This paper introduces an  extension of the matching/BSTS method for impact estimation to make it applicable for general conservation program impact estimation when multi-household data is available. This is accomplished by household matching/BSTS steps to obtain conservation estimates and then aggregating the results using a meta-regression step to aggregate the findings. A case study examining the impact of rebates for household turf removal on water consumption in multiple Californian water districts is conducted to illustrate the work flow of this method.
\end{abstract}

\end{frontmatter}

\section{Introduction}\label{introduction}

Many government agencies offer economic incentives to homeowners to reduce the environmental impact of their households. Common incentives are subsides for utility reducing appliances or rebates for conservation actions, such as removing conventional lawns in drought regions. Agencies that create these programs are faced with two questions during reviews: (1) how effective is the marketing of the conservation program to households, and (2) by how much does the incentivized conservation technology or activity reduce consumption of the target utility. Answering the first question is valuable for assessing prior marketing strategies and improving future outreach efforts, while answering the second is key to assessing the value of extending the program.

Internet marketing analysts also perform impact assessments to estimate how effectively an advertising campaign increases the so-called \emph{conversion rate} of the public, where online conversion rate is commonly quantified by clicks to an ad. \cite{google} propose a method for this type of analysis, where a treatment market exposed to the new advertising campaign is matched to another, similar control market. Then, the relationship between these markets prior to the campaign is modeled using Bayesian Structural Time Series (BSTS), and a counterfactual estimate of clicks per ad is produced during the campaign for the treatment market. The difference between the treatment (actual) and control (counterfactual) gives the estimated impact of the campaign. 

At the market level, internet marketers and conservation program managers are both interested in understanding if an outreach action was effective in stimulating engagement by individuals. Internet marketers are interested in clicks, while program managers are interested in participation in programs. Assuming that suitable benchmark markets can be found for conservation programs, it should be straightforward to apply the methodology introduced in \cite{google} to assess the effectiveness of a conservation program's outreach effort.

However, in this paper we will demonstrate how the matching and BSTS approach is particularly useful for estimating the impact of conservation programs on household consumption; in other words, for addressing the second question faced by agencies. To do so, the method proposed by \cite{google} is applied to individual households. Instead of estimating the number of clicks generated by a marketing campaign, the method is used to estimate how many units of a utility are saved by a household's participation in a conservation program. The method is further extended using a meta-regression step to systematically aggregate the estimates per household into general findings for informing policy.

The paper is developed as follows. First, in Section \ref{data}, a case study is introduced concerning rebate offers made to households for turf removal in California. This context will serve as the basis for summarizing the matching/BSTS methodology and its extension in Section \ref{methodology}. Then, in Section \ref{results}, the proposed impact estimation method is carried out and results are discussed. In Section~\ref{sec:discussion}, features of the method proposed in this paper and other approaches are discussed. Section \ref{conclusions} concludes.

\section{California turf rebate data}\label{data}
With outdoor landscaping representing approximately half of urban water usage, the Californian water management community has identifed outdoor water usage in general \cite{mayer2015}, and ornamental lawns specifically \cite{CUWCC2015} as a key opportunity in the larger effort to increase water conservation. Between July 2014 and April 2016, the Metropolitan Water District (MWD), the regional wholesaler of Colorado and Bay Delta water for Southern California, paid out \$270.7 million directly for turf rebates under its regional program and another \$15.1 million to supplement member agency spending on turf replacement. Metropolitan indirectly serves 6.1 million residential households across Southern California \cite{mwd2016}. In addition, millions in local retailer turf rebate supplements have been paid out (for example in Los Angeles, Long Beach, San Diego and Moulton Niguel).

The data used in this paper to illustrate conservation program impact estimation was provided by 3 water utilities: Moulton
Niguel Water District (MNWD), Irvine Ranch Water District (IRWD), and
Eastern Municipal Water District (EMWD). Each utility provided two data
sources. The first is a panel data set of monthly billed water usage and
customer characteristics identified by account and service point (water
meter) identifiers for single family households. The second is a data
set detailing participation in water efficiency rebate programs, of
which turf removals are the primary interest for this study.

These two data sets are merged, and any turf rebate instances tied to
accounts that appear more than once are dropped to prevent over-counting. Observations with clearly extreme values for variables used in a modelling step were also identified and removed using visual inspection and simple multivariate robust regression models.
The remaining accounts are then further restricted to those that have at least
two years of data (24 observations) in the pre-rebate period and one
year of data (12 observations) in the post-rebate period. The pre- and
post-rebate periods are determined relative to the month that the
post-rebate inspection was performed. Finally, the water districts make
use of default values in cases where the actual value is unknown.
Some districts substitute default values for irrigable area when actual values are not known. Customers with default values were dropped in cases where this was obvious due to bunching of many customers at the same value of irrigable area. 

After cleaning, the California Turf Rebate Dataset (henceforth, CTRD) that serves as the working dataset in this paper contains 545 households that received either traditional or synthetic turf rebates. The variables are defined as follows:

\begin{itemize}
\item
  \texttt{Customer ID}: unique identifier for each household.
\item
  \texttt{Month} and \texttt{Year}: the month and year of the water
  bill.
\item
  \texttt{HH Size}: number of permanent residents at the
  property.
\item
  \texttt{Irr Area Sf} and
  \texttt{Rebate Quantity}: the square feet of irrigable area
  and square feet of turf removed during rebate, respectively.
\item
  \texttt{Rebate Area Ratio}: the proportion of turf area removed,
  calculated as $\frac{ \texttt{Rebate Quantity} }{\texttt{Irr Area Sf} }$.
\item
  \texttt{Evapotranspiration}: The reference evapotranspiration, $ET_0$, in inches.
\item
  \texttt{Population density 2015}: population density of the zip code where a household is located. 
\item
	\texttt{Pre-turf removal efficiency}: a measure of a household's deviation from that household's allocated water budget. A water budget is the quantity of water calculated by the district as adequate for a given household. With $p$ denoting the last time point prior to turf removal for treatment household, and that household's consumption denoted as $tr$, the mean log ratio of used water over budgeted water at each period prior to turf removal is:
	\begin{equation}
	\frac{1}{p}\sum_{t = 1}^p log \frac{tr_t+1}{\text{[budget tr]}_{t}+1},
	\end{equation}
	where 1 is added to the usage and budget to avoid dividing by zero. Values below 1 indicate that the household is a water conserver prior to turf removal, while values above 1 indicate the opposite. This variable gives an indication of the water usage behavior of the household prior to turf removal.
	
\end{itemize}

\section{Methodology}\label{methodology}
The matching/BSTS methodology outlined by \cite{google} focuses on measuring the impact of a discrete marketing event. In the conservation program context, analogous example events are a rebate offer, a change in pricing structure or an advertising campaign; but the focus of this paper is on the non-analogous application of measuring the impact of a physical change to a household on utility usage. The aim of the matching/BSTS analysis is to quantify the event's impact on a response metric of interest (e.g., household consumption of a utility). The causal impact of a conservation action, or treatment, is the difference between the observed value of the response and the (unobserved) value that would have been obtained under an alternative set of circumstances. Usually the alternative circumstances of interest are those that would have occurred had no marketing or monetary offering event occurred, or when no physical change to a household occurred. This case is frequently referred to as the \emph{counterfactual}.

The innovation of the matching/BSTS methodology for causal impact estimation is the construction of a counterfactual based on two steps. First, one or more time series from a pool of candidates are classified by a matching algorithm as similar enough to the treated time series in the pre-treatment period that they can be used to infer trends that would influence the behavior in the treated series post-treatment. Second, using a BSTS, the relationship between the matched series and the treated series is modeled pre-treatment and used to predict the treated series post-treatment. This post-treatment prediction is the matching/BSTS counterfactual of the treated series under the scenario where no treatment was applied. The difference between the actual and predicted consumption of the treatment series is considered the impact of the treatment.

Many utilities collect data on utility consumption and conservation program participation at the consumer level. One can apply the methodology just described to a consumer that has participated in a conservation program. In this case, the methodology detailed in \cite{google} and used in market impact analysis by online marketers is suitable. However, unlike a single market, a single consumer is rarely of much interest. Rather, it is desirable to apply this method to many users to gain deeper insights into broad consumer behaviors, and the behaviors of subgroups of consumers. One way to obtain an aggregated estimate of consumer behavior is to perform a meta-regression.

Meta-analysis is the practice of pooling estimates from multiple studies on the same effect to expose general properties of the effect. This approach is popular in the medical statistics literature, where it is common to pool estimates from studies of a medical phenomenon or treatments conducted in different regions or through different programs because there is typically reason to believe that a study in a single location will not completely explore the characterstics of the phenomenon or treatment due to cultural or institutional confounding. In the CTRD, estimates of turf removal impact on water consumption are obtained for 545 households that filed rebates. Each estimate of the rebate impact might be thought of as an estimate for the effect of a medical treatment obtained by a single study.

The steps below describe the matching/BSTS process and its extension in more detail in the context of the turf rebate case study data. Given \(N =\) 545 treatment accounts which participated in a turf removal rebate and are examined in this study:

\begin{enumerate}
\def\labelenumi{\arabic{enumi}.}
\item
  Each treatment account \(tr_i, i \in 1 \dots N\) which has participated in
  a turf rebate is matched with a set of control accounts
  $C_k = \{c_k^j\}, j \in 1, \dots, 6, k \in 1, \dots, K$ from the same zip code which did not
  participate in a turf rebate. These control accounts are chosen by how
  similar their historical usage patterns are to the usage patterns of
  the treatment account \(tr_i\), based on a weighted combination of
  their Pearson correlation and their warping distance.
\item
  After the \(c_k\) have been chosen, we fit a BSTS model and use it to estimate the monthly impact of turf removal
  on water savings. The BSTS model uses the
  water usage patterns of the control accounts to create a synthetic
  control corresponding to the expected water usage of \(tr_i\) if there
  had been no turf removal. The predicted usage in the post-rebate
  period is then subtracted from observed usage to obtain a monthly
  water savings estimate for \(tr_i\).
\item
  After water savings estimates have been calculated for each treatment
  account, the last step is to obtain an overall summary estimate. This
  is done with a meta-regression approach that uses the estimates and
  a measure of estimate accuracy from each treatment account as the inputs into a random
  effects model.
\end{enumerate}

For the sake of brevity, this workflow will be referred to henceforth as M123; (1) Matching, (2) Modeling, (1) Meta. As shorthand, steps will be referred to as, for example, M1 for matching, or M13 for matching and meta. The first two steps are implemented into a workflow by the
\texttt{MarketMatching} package.
\footnote{The code was modified and is available at https://github.com/christophertull/MarketMatching/tree/usability-improvements}

\subsection{Choosing Control Accounts}\label{choosing-control-accounts}

The first step in obtaining an estimate of the turf removal impact for
account \(tr_i\) is to find accounts that did not remove their turf that
show similar behavior to \(tr_i\). Candidate accounts were identified by
choosing controls from within the same zip code as \(tr_i\). Within each
zip code there may still be thousands of possible controls. These
remaining possibilities are ranked by how similar their historical water
usage patterns are to the historical usage of \(tr_i\).

Account matching is often based on variables like property size,
property value, or education levels. However, the importance of
environmental attitudes, for example arising from public awareness
actions and social change has been shown to influence water consumption \cite{hollis2016}. The difficulty of incorporating these and other
difficult-to-quantify factors driving household water usage, and the
fairly stable water consumption patterns observed by most households,
make matching based on water consumption patterns attractive. The
premise of using historically predictive relationships between accounts
to perform counterfactual analysis in this fashion has been advocated by, for example, \cite{abadie2010} and \cite{google}.

Let \(tr\) and \(c\) be a treatment and control time series with \(p\)
observations each for which a similarity ranking is desired. This
similarity ranking is specified as a weighted composite of two other
similarity measures. The first is the Pearson correlation:

\begin{equation}
 \rho(tr,c)=\frac{\sum_{t=1}^p(tr_t-\bar{tr})(c_t-\bar{c})}{\sqrt{\sum_{t=1}^p(tr_t-\bar{tr})^2}\sqrt{\sum_{t=1}^p(c_t-\bar{c})^2}} \nonumber
\end{equation}

The second ranks them according to their dynamic time warping (DTW)
distance from \(tr_i\). To compute the warping distance between two time
series, we must identify the warping curve
\(\phi(t)=(\phi_{tr}(t),\phi_c(t))\) that has the minimum warping
distance,

\begin{equation}
D(tr,c) = \sum_{t=1}^p d(\phi_{tr}(t),\phi_c(t))m_\phi(t), \nonumber
\end{equation}
where \(d(\phi_{tr}(t),\phi_c(t))\) is the local of the points at time
\(t\) after they have been remapped by the warping functions
\(\phi_{tr}(t)\) and \(\phi_c(t)\), and \(m_\phi(t)\) is a per-step weight
that control the slope of the warping curve. The calculation of the DTW
distance is done using the \texttt{dtw} package in \texttt{R}. For details about the
package and about dynamic time warping see \cite{giorgino2009}.

Let the vector \(\pmb r\) denote the similarity scores for \(K\)
candidate control accounts \(c_k\) with respect to
\(tr_i\), where the \(k\)th element of  \(\pmb{r}\) is given by:
\begin{equation}
\label{lab:simMetric}
r_k = (1 - \alpha) \rho(tr_i,c_k) + \alpha D(tr_i,c_k); \;\alpha \in [0,1].
\end{equation}
Then, the control households corresponding to the
first \(m\) largest values in \(\pmb{r}\) are used as controls for
\(tr\) in the structural time series model for \(tr\) discussed
in the next section.

\subsection{Estimating Water Savings}\label{estimating-water-savings}

A widely used approach for estimating the impact of
interventions, like rebate offerings, is differences-in-differences.
Taking this approach in the turf removal context, the estimated
impact of turf removal on water savings is the difference between water
usage when turf was removed, and the amount of water that would have
been used if no turf had been removed \cite{bamezai1995}.

To accurately estimate the reduction in water usage due to turf removal,
a model for the counterfactual case needs to account for other variables
determining water usage. Water use is determined by a multitude of
factors, such as weather, household size, social perspectives on water usage,
and turf removal. Covariates like weather and household size are measured by
agencies and are straightforward to account for as covariates in a model.

This leaves the matter of accounting for dynamic behavioral patterns.
Recognizing the need to address this aspect of water use, \cite{hollis2016}
examined how variables measuring media factors, such as advertising volume,
explain water use patterns. The inclusion of media presence explicitly
in a usage model is desirable, but two issues that arise with this
approach are properly quantifying media presence and accounting for the
different levels of exposure experienced by water users.

Another way to account for dynamic behavior and classical covariates simultaneously is to explicitly model the
counterfactual of a time series observed both before and after the
rebate and use the resulting model to construct a synthetic control (cf.
\cite{abadie2010}). The approach of \cite{google} is to construct a synthetic control by combining three
sources of information using a state-space time-series model: (1) behavior of the pre-treatment target series (2) other time series
that were predictive of the target series before the turf removal, and (3) in a Bayesian framework, prior
knowledge about the model parameters, from earlier studies.

We will use static regression coefficients in our Bayesian structural
time series model, which assumes that the linear usage relationship
between the controls and the counterfactual expected usage for customers
who did remove turf from their lawn remains fixed even after the turf is
removed. Furthermore, we will allow for a local linear trend. For a treatment time
series \(\emph{\pmb{tr}}\), this model has the form:

\begin{eqnarray}
tr_t=\underbrace{\mu_t}_\text{level}+\underbrace{Z_t}_\text{regression}+\varepsilon_{t}, \nonumber\\ \nonumber
Z_t = \beta'\pmb{x},\\ \nonumber
\underbrace{\mu_{t+1} = \mu_t + \delta_t + \eta_{\mu,t}}_\text{random walk and trend},\\ \nonumber
\underbrace{\delta_{t+1} = \delta_t + \eta_{\delta,t}}_\text{random walk for trend},
\end{eqnarray}
where \(\varepsilon_t \sim \mathcal{N}(0,\sigma^2_t)\),
\(\eta_{\mu,t}\sim \mathcal{N}(0,\sigma^2_{\mu,t})\) and
\(\eta_{\delta,t}\sim \mathcal{N}(0,\sigma^2_{\delta,t})\). The
regression component, \(Z_t\) captures the static linear relationship
between the control series and the treatment series, while the level
component \(\mu_t\) captures local linear trends, enabling the model to
react to unobserved sources of variability the control and treatment
series are exposed to. Note that while the vector $\pmb x$ can contain covariates besides an intercept, classic covariates, such as household size are not explicitly included in the model implemented in this paper (nor in the model implemented by \cite{google}), but are implicitly captured since the modelled trend is conditional on them.

By placing a spike-and-slab prior on the set of regression coefficients,
and by allowing the model to average over the set of controls, it is
possible to choose from many candidate controls \cite{george1997}. To combine information about \emph{\pmb{tr}} and the
controls, the posterior distribution of the counterfactual time series
is computed given the values of \emph{\pmb{tr}} in the pre-intervention
period, along with the values of the controls in the post-intervention
period. Given a predicted and observed water use \(\hat{tr}_t\) and
\(tr_t\), the difference \(tr_t-\widehat{tr}_t\) yields a semi-parametric
Bayesian posterior distribution for the water savings attributable to
the turf removal at time $t$, which can be used to obtain credible intervals. When water is considered saved, the value of $tr_t-\widehat{tr}_t$ is negative since less water was consumed than expected given the counterfactual water consumption $\widehat{tr}_{t}$. We
take these estimates and adjust them from CCF/hundred cubic feet to gallons saved per square foot to obtain estimated monthly gallons saved per square foot of turf removed, calculated as:
\begin{equation}
\label{eq:savings}
\widehat{gpsf}_t = \frac{748.052 \times (tr_{t} - \widehat{tr}_{t}) }{ \texttt{Rebate Quantity}} \nonumber
\end{equation}


The structural time series model was fitted using the \texttt{CausalImpact}
package provided by Google for estimating the effectiveness of marketing
campaigns \cite{google}. A number of differences exist between
the Google marketing context described in \cite{google}, for
which this approach was originally proposed, and the turf removal rebate
context. Firstly, Google is able to assess the impact of the marketing
campaign in terms of participation using this method, where
participation is measured in number of clicks, because they have data on
number of clicks prior to the campaign. It is in their interest to
distinguish how many clicks after the start of the campaign were driven
by the campaign, as opposed to organic. In contrast, prior to the rebate
programs, the water districts did not track turf removal. The number of
filed rebate claims before the start of the rebate programs is zero, so the rebates do not definitively show that turf removal has increased, unlike advertising clicks, which are monitored before and after an marketing action.

Another difference is that in the marketing context, the impact to
estimate is the number of clicks generated as a consequence a marketing
campaign, where a marketing campaign is either active or is not. The
scale of the marketing campaign is not addressed. We could stop at
estimating an average savings of a household that removes turf, but this neglects the
important relationship between how much water use is reduced and the
amount of the turf removed. To account for this, the estimated savings
are divided by the square feet of turf removed, as calculated by utility
staff in a post-rebate inspection (Equation \ref{eq:savings}). This allows for a normalized measure
of rebate impact in terms of gallons per square foot of turf removed.
Additionally, variables to quantify the magnitude of the turf removal
are included in the meta-model in the final step.

An added complexity in this study is that in place of a single treatment
cohort, or perhaps a few, hundreds of customers participated in the
rebate program. The approach proposed in \cite{google} stops
at providing impact estimates on a single time series at a time. To
obtain a broad overview of the impact of turf removal, it is desirable
to aggregate estimates from all of the customers. This issue and the inclusion of the amount of turf removed in
our framework will be addressed using a meta-regression approach. This meta-regression step will also allow us to explicitly extract the contribution of classical covariates, such as household size, whose influence is only implicitly modeled in the M2 step.

\subsubsection{Example of M12}\label{example}

Figure \ref{fig:example} below shows two examples of the process described above. Specifically, the output of the matching process is shown through charts of water usage over time for the treatment household and its six closest matches. The output of the BSTS model is given by showing the actual and predicted consumption for the two examples. The example households were chosen for their wildly different behavior patterns in the post-rebate period. One household appears to cease outdoor watering completely after their turf removal, causing their usage to stabilize at winter levels and achieving an estimated 66\% reduction in overall water use. The other example household shows a decrease in usage relative to its own past behavior, but shows no significant reduction compared to its similarly-behaving peers. This effect may be due to increased awareness of the California drought and the mandatory restrictions put in place in April 2015. Thus the water savings would be attributed to behavioral change among households in the region but not directly to the removal of turf.

\begin{figure}
	\caption{The first row shows the expected and observed usage
		patterns for two participating rebate accounts, where the difference
		between expected and observed after removal (dashed) is the estimated savings. The account on the left shows a visible reduction in usage compared to the counterfactual, while the right side has more ambiguous results. The bottom row shows the raw time series of
		water usage for the treatment and corresponding matched controls.}
	\centering
	\includegraphics[width=0.95\textwidth]{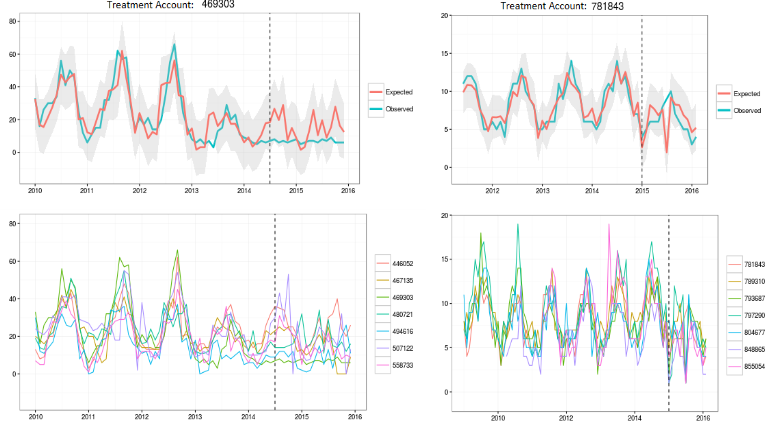}
	\label{fig:example}
\end{figure}


\subsubsection{Parameter selection}

A number of parameters must be chosen when applying the matching
procedure and BSTS model. We assessed these in terms of their impact on
the mean water savings estimates obtained from the BSTS models.

A sensitivity analysis was performed to determine the effect of
parameter choices at the matching stage on final estimates of water
savings. Specifically, a random sample of 150 accounts were rerun under all combinations of the different parameter
configurations visible in Table \ref{table:sensitivity}. While these are not the only parameters in the model, they are three of the ones most likely to impact the water savings estimates because they directly impact the choice of control accounts.

\begin{table}[]
\centering
\label{table:sensitivity}
\begin{tabular}{ll}
Parameter                               & Values                     \\\hline
WARPING LIMIT                           & 0, 1                       \\
DTW EMPHASIS                            & 0, 0.25, 0.5, 0.75, 1      \\ 
NUMBER OF MATCHES 						& 6, 12 					\\ 
\end{tabular}
\caption{Parameter Values tested in sensitivity
analysis.}
\end{table}

In the BSTS model, the value for \(\sigma^2_{\mu,t}\) in the local linear
trend must also be selected.  This is the local level standard deviation which controls the prior standard deviation of the local linear trend submodel. The local level term modifies how adaptable the model is to short term changes, and its standard deviation is important because it affects the breadth of the
posterior intervals. \cite{google} recommend that the
value of 0.01 can be used when the relationship between the controls and
the treatment is strong enough to obtain an informative model. The
authors indicate that this is more likely when many control candidates
are available. The water usage data set contains a large pool of control
candidates, and matching results are typically strong (high values in the \emph{\pmb r} vector, and visually convincing matches), so 0.01 is taken.

After calculating savings estimates under each parameter set, the mean of estimated savings for the sample under each parameter set was calculated. This gives an idea of how sensitive the matching process is to changes in the parameters. These estimates are visible
below in Figure \ref{fig:sensitivity}.

\subsubsection{Computation}
A draw-back of the M12 approach in conservation analyses is that the procedure needs to be applied to hundreds or thousands of households, but it is computationally heavy. This burden can be managed by cutting the time of individual operations or parallelizing them. 

\begin{figure}[h]
\centering
\includegraphics[width=80mm]{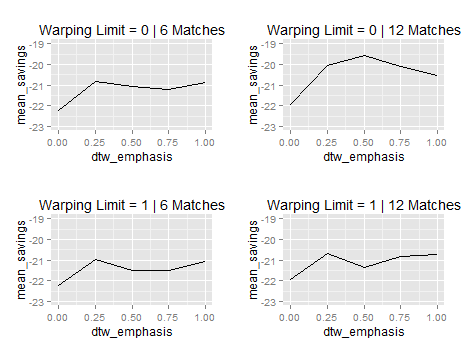}
\caption{The charts display the sensitivity of the meta-estimate results
under various values of the DTW EMPHASIS parameter. Each chart in turn
uses a different warping limit or number of control account matches.}
\label{fig:sensitivity}
\end{figure}

The most computationally expensive step in the M12 steps is the search for suitable matches. A major source of computational cost of this step is the DTW calculations used to obtain $D$ in Eq.~\ref{lab:simMetric}. Figure~\ref{fig:sensitivity} shows that there is some sensitivity in the meta-regression estimates (the analytical end-point of this methodology) as a function of $\alpha$, the DTW emphasis parameter. However, if the analyst reaches the conclusion that matching results are not sensitive to the emphasis on the DTW component, then a choice can be made to set $\alpha$ in Equation~\ref{lab:simMetric} to 0. This eliminates the DTW calculations and substantially increases the computation speed of the procedure. 

A second way to improve operation time is to parallelize the M12 steps. This topic is not addressed in~\cite{google}, possibly because the authors are addressing the normal situation in market impact analysis where an ad campaign is applied to one or a small number of markets and matches need to be found from perhaps a few hundred markets. Such a small number of M12 operations does not justify parallelization since even a single core can complete the task quickly. In contrast, when assessing the impact of a conservation action on hundreds or thousands of households, with thousands of candidates to match against, it becomes highly beneficial to take advantage of the distinctness of the household-level M12 calculations. 

Two levels of the M12 steps can be parallelized. In the simpler approach, parallelization is performed at the participating household level according to the following steps:

\textbf{Parallelization scheme 1}
\begin{enumerate}
	\item Select a participating household $i$ and candidates for matching (based on geographic proximity, zip code, etc).
	\item Distribute the data for the selected participating $i$ and matching candidate households to a core.
	\item perform the M12 steps and return the estimated conservation impact $\widehat{\pmb{gpsf}}_i$ from the core for each time point in the post-rebate period.
	\item Perform prior steps for all participating households as cores become available.
\end{enumerate}

When the number of cores available is large and the number of candidates to match is extremely large, it may be advantageous to further distribute the computing as follows:

\textbf{Parallelization scheme 2}
\begin{enumerate}
	\item Determine a number of cores $C$ that will be devoted to matching for a household $i$.
	\item Select a participating household $i$
	\item Select $C$ subsets of $\lceil K/C \rceil$ matching candidates (based on geographic proximity, zip code, etc), such that for any two subsets, $C_c \cap C_{\neg c}=\varnothing$, or nearly so, where only slight overlap occurs due to rounding.
	\item Distribute the data for the selected participating $i$ and matching candidate households to core $c$.
	\item Perform the matching step and return the similarity index for each candiate in $C_c$.
	\item Combine the similarity indexes to obtain the vector \(\pmb{r}_i\) and select the matching households. 
	\item Run the BSTS step and estimate the conservation impact series $\widehat{\pmb{gpsf}}_i$ on an available core.
	\item Perform prior steps for all participating households as cores become available.
\end{enumerate}

Table \ref{table:param-choices} shows the values of the matching procedure parameters based on
the results from the sensitivity analysis, as well as required minimum
observation period lengths and matching pool sizes.

\begin{table}[]
\centering
\caption{Key Parameter choices in the modeling process.}
\label{table:param-choices}
\begin{tabular}{p{3cm}p{6mm}p{4cm}} \hline
Parameter               & Value & Description                                                                                \\ \hline
Min. Months Post-Period & 12    & Require at least 12 months since the rebate took place.                                    \\
Min. Months Pre-Period  & 24    & Require at least 24 months before the rebate for accurate matching.                        \\
Zip Sample Size         & 500   & Randomly sample a maximum of 500 control accounts within the zip code as possible matches. \\
Min. Matching Series    & 100   & Require a pool of at least 100 possible matches within the zip code.                       \\
Warping Limit           & 1     & The size of the Sakoe-Chiba band limiting how much the time series are allowed to warp.    \\
DTW Emphasis            & 0.7   & Controls the trade-off between the DTW distance and Pearson correlation.                   \\
Number of Matches       & 6     & The number of control accounts to match with and pass into the STS model. \\
\hline                 
\end{tabular}
\end{table}

\subsection{Meta-regression on the savings estimates}\label{combining-the-estimates}

Monthly estimated water savings attributable to turf removal are
obtained from each of the BSTS models, yielding a total of 10759
impact estimates for 545 households. Furthermore, a credible interval
can be calculated for each of these estimates. In addition to considering this large collection of estimates in their own right, we can also model them to reveal general patterns linking estimated savings to known household characteristics. Doing so allows us to make a step from the household BSTS models, which do not distinguish between behavioral trends and the influence of covariates, to general insights into drivers of water savings that are classically of interest.

The aggregation of the household savings estimates can be accomplished with meta-regression, which allows us to include covariates that we expect to drive water consumption, such as household size, and to account for variability between the households. More specifically, we will use a mixed-effects model for $\widehat{\text{gpsf}}_{i,t}$, the savings of the $i$th household at time $t$,:
\begin{equation} \label{mixedFit}
\widehat{\text{gpsf}}_{i,t} = \beta_0 + \beta_{1}x_{i,1,t}+\dots+ \beta_{p} x_{i,p,t}+ u_i+\varepsilon_{i,t}, \nonumber
\end{equation}
where $\beta_1\dots\beta_p$ are coefficients for selected fixed effects, $u_i\sim \mathcal{N}(0,\tau^2)$ is a random intercept per household and $\varepsilon_{i,t}\sim \mathcal{N}(0,v_{i,t})$ is a random error term. One of the underlying assumptions of many modeling techniques is that each observation is measured with equal precision. However, in this context, we have an estimate of the precision for each water usage prediction, courtesy of the highest posterior density credible intervals returned by the BSTS. Each credible interval can be characterized by an estimated variance $s_{i,t}$, which can be used to compute weights for each observation: $w_i=1/(s_{i,t})$. By applying these weights to the observations in the meta-regression, observations for which the BSTS model was able to provide good predictions are given the most influence. The meta-regression is performed using the \texttt{metafor} package \cite{metafor}. The sources in \cite{metafor} also provide background on the respective methods used in mata-analysis.

\section{Case study: Analyzing the CTRD}\label{results}
Having outlined the steps for performing a conservation impact analysis using the M123 approach, we return to the CTRD. Our goal is to understand how turf removal and other variables influence household water consumption.

After matching and performing the BSTS step, we obtain an estimated change in water consumption per household that participated in the rebate program. This estimate is a quantification of the total difference between the expected and realized water consumption at each time period. To better understand how turf removal influences water usage, the meta-regression step can be used to decompose this change in consumption into factors of interest, such as the magnitude of turf removal and household size.

The meta-regression we conduct to aggregate the results from the Bayesian
STS model estimates of the water savings from the \(i\)th turf-removing
household at time \(t\) is a mixed effects model with the following
fixed effects structure:

\begin{equation} \label{eq1}
\begin{split}
\text{$\mu$ gpsf}_{i,t} & = \alpha_i + \beta_0 + \beta_1\times \text{HH Size}_{i,t} \\
&+\beta_2\times \text{Pre-turf removal efficiency}_{i,t} \\
&+\beta_3\times ln(\text{Rebate Quantity}_{i,t})\\
&+\beta_4\times ln(\text{Rebate Area Ratio}_{i,t}) \\
&+\beta_5\times \text{Household income}_{i,t} \\
&+\beta_6\times \text{Evapotranspiration}_{i,t} \\
&+\beta_7\times \text{Population density}_{i,t} \\
&+ \beta_8\times sin(2\pi/12 Month_{i,t}) \\ 
&+ \beta_9\times cos(2\pi/12 Month_{i,t}) \\
&+\beta_{10}\times sin(4\pi/12 Month_{i,t}) \\
&+ \beta_{11}\times cos(4\pi/12 Month_{i,t}) \\
&+ u_i+\varepsilon_{i,t}, \\ \nonumber
\end{split}
\end{equation}
where $\text{$\mu$ gpsf}$ is the monthly savings in gallons per square foot. The trigonometric terms in the model account for seasonality and general time trends in savings not captured by the BSTS model in the pre-removal period. Month, in
this model, is a unique number for each of the months in the study and
runs from \(1,\dots,51\).

Table \ref{tab:modelOutput} contains the fixed effects estimates of the
fitted model. Negative values indicate that as the value of the variable increases, water is increasingly saved. The response variable is estimated water savings from households that removed turf, and the intercept gives a baseline value of around 17 gallons saved per square foot per month.  Household size is not significant, though the direction is positive. Pre-removal efficiency has a strong, negative effect. This result indicates that households which tend towards excessive consumption prior to turf-removal save more per square foot. The log of rebate quantity is positive and significant, indicating that there are diminishing returns per square foot removed. The variable rebate area ratio displays similar behavior.

Above the single household level, zip codes with higher median household incomes tend to save less water per square foot of turf removed. This may be because these households practice less conservative gardening to start with. Population is only significant at the 10\% level, but the direction is that areas with denser population save less per square foot removed. Finally, higher ET values result in higher savings, which is because when ET is high, yards that would normally be watered heavily, are not watered at all.

The trigonometric effects used by the model to capture general time trends in water savings are significant. The fact that time terms are significant indicates that the BSTS model does not completely anticipate how monthly developments influence water usage in houses that removed turf relative to their controls. An explanation for this is that during the training period the BSTS models a relationship between control and treatment houses that have lawns which are generally fixed in size. After turf removal, the relationship between these houses changes; water consumption by houses that remove turf reacts differently in the dry months than before removal because less watering is required. This explanation also applies to the ET variable, which explains higher savings than the BSTS models expect because the water savings benefits of turf removal are not available when fitting the BSTS model since that is pre-removal. 

In preliminary specifications of the model, dummy variables were included for the different districts in the data set. An attractive feature of a multi-district dataset is the potential that different implementation regimes will create conditions for a natural experiment. However, the district effects were insignificant. This is potentially due to the similarity of the districts in this study, which are geographically nearby in Southern California. A greater variety of districts in the dataset could offer the potential for more insights.

\setlength{\tabcolsep}{2pt}

\begin{table}[h!]
\centering
\caption{Fixed effect estimates for the meta-model of turf removal water savings.} 
\label{tab:modelOutput}
\begin{tabular}{lrrrr}
  \hline
Variable&Estimate&SE&t-stat&p-value\\
Intercept&-17.66&5.07&-3.48&0\\
HH Size&0.08&0.05&1.5&0.13\\
Pre-Removal Efficiency&-4.36&1.06&-4.1&0\\
ln(Rebate Quantity)&0.55&0.23&2.36&0.02\\
ln(Rebate Area Ratio)&0.81&0.27&3.02&0\\
ln(med. HH Income)&1.06&0.41&2.6&0.01\\
Population Density&1E$^{-4}$&1E$^{-4}$&1.65&0.1\\
ET&-0.08&0.03&-2.86&0\\
Month Sin 2&0.43&0.03&13.14&0\\
Month Cos 2&0.27&0.07&3.98&0\\
Month Sin 4&0.09&0.03&3.45&0\\
Month Cos 4&0.02&0.03&0.7&0.48\\
   \hline
\end{tabular}
\end{table}

\begin{figure}[h]
\centering
\includegraphics[width=80mm]{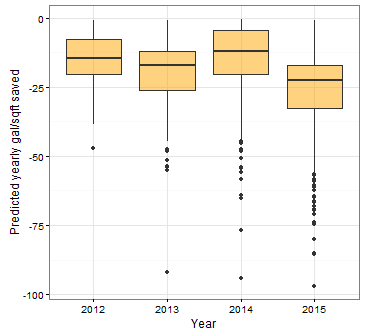}
\caption{Average yearly savings for each household over 5 years.}
\label{fig:yearAvgSavings}
\end{figure}

The first post-modelling analysis we conduct is a comparison of average household savings by year. 
We do this for the sample in this study by
using the model to predict the household savings given their moderator
variables. Predicted savings are then grouped by household and year and
averaged. The resulting savings estimates give an impression of the
distributions of savings outcomes that would be expected by an analyst
or policy-maker on this population (Figure~\ref{fig:yearAvgSavings}). We see that annual savings were
about 20 gallons per square foot. However, by aggregating the monthly
savings to an annual level, we lose important details about the savings
patterns.

\begin{figure*}
\centering
\includegraphics[width=0.8\textwidth]{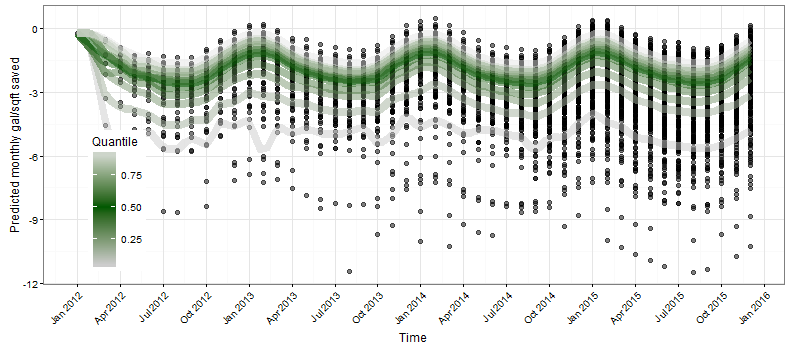}
\caption{Predicted monthly savings for each household in the data set. The dark green line corresponds to median savings. Seasonal variation leads to swings in average savings from $-1.5$ to $-2.7$ gallons per square foot.}
\label{fig:monthly_predictions}
\end{figure*}

A more nuanced approach is to use the model to predict the monthly household
savings. Overlaying the predictions are their quantiles, ranging from 5\% to
95\%. The savings pattern illustrated in Figure \ref{fig:monthly_predictions} is highly intuitive. The highest savings are in the months of July, August and
September, reaching a monthly average reduction of 2.7 gallons per square
foot. During the months of January, February, and March,
the reduction is much smaller but still valuable at -1.5 gallons per
square foot. Additionally, the plot depicts the skewness of the savings distribution. Turf removal nearly always results in some savings, and savings for most households that save more than the average fall within a bandwidth of one gallon more than the average, with a small proportion of extreme savers. Examining the predictions against their quantiles is similar to the practice advocated by~\cite{Riley2011} of examining prediction intervals in random-effects meta-analysis in order to have a more complete overview of the range of potential treatment outcomes.

\section{Discussion}
\label{sec:discussion}
The M123 results for the CTRD quantify the contribution of turf removal to water savings and elaborate on how these savings are realized. However, M123 is not the only approach for obtaining these estimates.

\subsection{Time-Series vs.~Traditional Matching}
\label{time-series-matching-vs.traditional-matching}

One remaining question of interest is whether time series matching on
historical usage produces comparable results to traditional matching on
static attributes. In order to address this question, the mean distance
from each treatment account to its matched control accounts was compared
to the mean distance from each treatment to its potential controls that
were not matched.

\begin{figure}[h]
\centering
\includegraphics[width=85mm]{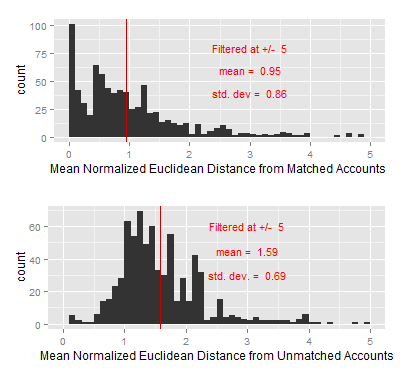}
\caption{The top histogram shows the distribution of mean distances
between treatments and their matched controls with similar historical
usage. The bottom shows the distances between treatments and the
unmatched accounts with more dissimilar usage patterns. On average,
accounts with similar usage tend to be more similar in household size
and irrigable area than those with very different usage patterns.}
\label{fig:euclidean}
\end{figure}

Distance was calculated by standardizing the covariates for household
size and irrigable area within each zip code and customer class. The
mean Euclidean distance was then calculated between the treatment and
each of the matched and unmatched groups. The results of this
calculation are visible in Figure \ref{fig:euclidean}. One can see that matching on usage
patterns tends to result, on average, in matches that are also similar
in their household size and irrigable area. However, this was not
universally true and manual inspection revealed a large variation even
among the matched control accounts. This aligns with the intuition that
static covariates do not capture all aspects of water usage, and that
dissimilar accounts may have very similar water usage patterns.

\subsection{BSTS vs. mixed model with dummies}
The M123 approach uses a BSTS model to obtain $\widehat{gspf}$. Consider instead an approach based on a linear mixed model with dummy variables for post-rebate observations
\begin{equation}
use_{i,t} = \beta_0+\beta_1x_{1,i,t}+\dots+\beta_p x_{p,i,t}+\delta d_{i,t}+u_i+\varepsilon_{i,t},
\end{equation}
where $d_{i,t}$ is a dummy variable indicating if a household has removed turf, and $\delta$ is a coefficient quantifying this effect. Given that we know the impact of turf rebates is conditional on climate factors, we could also expand the list of dummy variables to include interactions with season and ET. Along this line, many papers concerned with modeling water consumption use mixed models containing covariates thought to correlate with water usage and random intercepts to control for between-consumer variation (\cite{fielding2013, harlan2013, mitchell2013, olmstead2007, EndterStrata})
 
Although both M123 and the mixed model approach both include individual unit information in the model, they obtain estimates for different questions. The M123 estimates to what extent a participating household's consumption is different from what it would have consumed if it had not participated. A mixed model estimates how consumption by participating households is different from non-participating households, by means of the dummy variable. The M12 approach is interesting if one is concerned with selection bias, wherein households that participate in conservation programs are likely to consume less water to begin with. The mixed model approach can not disentangle the a household's concern for water use and water use reduction incurred by technical savings from a modification, such as turf removal. In contrast, the M12 approach makes some progress in this direction by modeling household water use idiosyncrasies in the pre-treatment phase that may be accounted for by unmeasurable preferences, such as concern for water use.

\subsection{Improvement with quantile regression}
Figure~\ref{fig:monthly_predictions} gives a broader overview of the impact of turf removal than point estimates for a model of the mean, such as a table of coefficient estimates would. Ultimately, however, Figure~\ref{fig:monthly_predictions} is a detailed view of the prediction intervals generated by the model of the mean shown in Table~\ref{tab:modelOutput}. It is of interest to understand better how the behavior of households in different water consumption brackets relate to descriptive covariates. Quantile regression provides one tool for exploring these relationships \cite{quantileRef2005}. Employing a form of hierarchical spline models to estimate conditional quantiles, \cite{hendricks1992}, explains that in studying electricity consumption in Chicago households, that the relationships between consumption and covariates differ across consumption quantiles. The methodology applied by the authors bears some resemblance to the M123 method in that models are fitted per household and the results are aggregated up to general results. As the authors state, this ``hierarchical'' linear models approach was introduced by~\cite{lindley1972} and has been widely applied.

The M123 method differs from~\cite{lindley1972, hendricks1992} and others in the methods used in the M12 steps. Regarding the M3 step, an option which post-dates~\cite{hendricks1992} and is particularly suitable in concept is quantile regression with random effects \cite{quantileLongitudinal, fastQuantile2013}. In application, functions from the \texttt{rqpd} package \cite{rqpd} can be applied in the M3 step to perform the meta-regression. This approach was tested on the CTRD with some promising results, but the computational speed makes it prohibitively slow to apply to datasets of this size. Applying it to smaller datasets would be feasible, and there are prospects that methods such as \cite{fastQuantile2013} will allow for more rapid computation of these models in the near future.

\section{Conclusions}\label{conclusions}

The M123 methodology enables estimation of conservation program impacts using contextual customer attribute data of interest to policy makers and characterizations of usage dynamics based on observed household consumption. This is in contrast to approaches which rely purely on extensive lists of covariates that are at best proxies for consumption behavior. 

In the CTRD case study the impact estimate is 24.6 gallons saved per year per square foot of turf removed. Bootstrapped standard errors of
those predicted water savings are 0.11 gallons per year per square foot
of turf removed. Those water savings are stable across district and vary
sinusoidally over time highlighting the structural water savings of turf
market transformation for regional and statewide water reliability
initiatives. At \$2 paid per square foot turf removed and assuming a
hyperbolic discount rate of five percent over a landscape conversion
lifespan of thirty years, that translates into a present value of \$1422
plus or minus seven dollars per acre foot of water saved.

The M123 approach can be used to evaluate the water savings associated with other conservation rebates, other customer-level demand management interventions, and potentially other natural resource conservation programs in energy or natural gas. As the old adage goes, ``you cannot manage what you cannot measure'' and such rigorous impact evaluations can help public managers navigate the uncertainties of program design in conservation policy development.

Measuring savings at the household level, as in the CTRD, allows managers to
highlight promising technical solutions, like turf removal, but also target educational materials on efficient consumption practices to
customers that have seen dis-savings in their post-implementation period compared
to their expected counterfactual ultility use. In the CTRD case, this would mean, for example, sending educational materials to households that removed turf, but have saved less water than expected: perhaps because they are not aware of proper watering practices for their new native landscapes.

\section{Acknowledgements}\label{acknowledgements}

This research was funded through the California Data Collaborative. The
authors would like to thank the Moulton Niguel, Irvine Ranch, Eastern
Municipal, Las Virgenes Municipal, Santa Margarita, and Monte Vista
water districts, along with the Inland Empire Utilities Agency, the East
Bay Municipal Utility District, and the Metropolitan Water District of
Southern California for all of their support. The authors also thank Michael Hollis for his thoughtful comments and peer review.

\bibliography{biblio}

\end{document}